\documentclass[12pt]{article}
\usepackage{feynmf}
\def\la{\mathrel{\mathpalette\fun <}}

\def\fun#1#2{\lower3.6pt\vbox{\baselineskip0pt\lineskip.9pt
\ialign{$\mathsurround=0pt#1\hfil##\hfil$\crcr#2\crcr\sim\crcr}}}

\title{On the production of a lepton pair in the collision of
ultrarelativistic neutral particle with nonzero magnetic moment
with nuclei}
\author{I.V.~Gaidaenko, V.A.~Novikov, M.I.~Vysotsky \\
ITEP, Moscow, Russia}
\date{}
\begin{document}
\maketitle

\begin{abstract}

Explicit formulas which describe muon pair production in reaction $\gamma\nu
\to \mu^+ \mu^- \nu$ through neutrino magnetic moment are obtained and used
to derive in the leading approximation cross section of muon pair production
in $\nu N$-scattering due to neutrino magnetic moment. This cross section
appears to be proportional to $\log^4 E_{\nu}$. Comparison with
experimental data on tridents production provides an upper bound
$\mu_{\nu_{\mu}} < 4\cdot 10^{-8} \mu_B$, which is approximately two orders of
magnitude
weaker than that from $\nu_{\mu}e$ elastic scattering data.

\end{abstract}

\section{Introduction}

CHARM II and CCFR collaborations observed production of $\mu^+ \mu^-$-pairs in
$\nu_{\mu}N$-scattering data sample and soon more precise data 
 will be available \cite{1, 8, 9}. The data are well described by
the Standard Model through the diagram shown in Fig.1, where local 4-fermion
interaction represents $W$- and $Z$-exchanges. If one assumes that
$\nu_{\mu}$ has nonzero magnetic moment another mechanism of $\mu^+ \mu^-$
production shown in Fig.2 comes into play. The two mechanisms of $\mu^+
\mu^-$-production do not interfere, so one should simply add
probabilities, compare them with experimental data and get in this way the
bound on $\mu_{\nu_{\mu}}$.

Looking at the diagram in Fig.2 we observe that substituting  an 
electrically charged particle instead of neutrino one obtains the reaction of
the lepton pair production in the collision of two charged particles. This
reaction is well studied in the literature. The first paper was published in
1934 by L.D.Landau and E.M.Lifshitz \cite{2}, who observed that the cross
section of $e^+ e^-$ pair production in the collision of two nuclei grows with
the energy of colliding nuclei $E_N$ as $\ln^3 E_N$. In 60's, when a
physical program for $e^+ e^-$-colliders was prepared, the particle production
by  fusion diagrams was investigated in detail; for a review see \cite{3}.
In particular, subleading terms were calculated. However as far as we know,
the case when one of the colliding particles is neutral and   
emits  photon by  nonzero magnetic moment was not studied in
literature. So in this paper the corresponding cross section is calculated for
the first time. We observe that it increases with the energy  as
$\ln^4 E_{\nu}$, i.e. even faster than in the case of a  charged particle.

The calculation of the cross section consists of two parts. In Section 2
formulas for the spectrum and for the total cross section of $\mu^+ \mu^-$ pair
production in the reaction $\gamma\nu \to \mu^+ \mu^- \nu$ are presented in a 
leading logarithmic approximation. Section 3 contains the precise formulas
which describe $\gamma \nu \to \mu^+ \mu^- \nu$ reaction. In Section 4 with the
help of the equivalent photon ( Weiszacker-Williams) approximation the
spectrum of the muon pairs obtained in Section 2 is converted into that in
reaction $\nu N \to \nu N \mu^+ \mu^-$ and the total cross section of this
reaction in the leading  approximation 
is determined. In Section 5 the numerical estimates are compared with
the experimental data. In Conclusions we qualitatively compare the cases of
photoproduction on the charged and neutral particles and, in particular,
demonstrate how $\log^4 E$ dependence in a magnetic case converts to $\log^3 E$
dependence in an electric case. 

\section{Reaction \mbox{\boldmath$\gamma\nu\to\mu^+ \mu^- \nu$} in the 
leading logarithmic \\ approximation}

The amplitude of $\mu^+ \mu^-$-pair photoproduction on neutrino is described
by two diagrams shown in Fig.3. The photon emission by neutrino is described by
the following vertex:
\begin{equation}
V = \mu_{\nu}\bar{\nu}\sigma_{\alpha\beta} \nu A_{\alpha} q_{\beta} \;\; ,
\sigma_{\alpha\beta} =
\frac{1}{2}(\gamma_{\alpha}\gamma_{\beta}-\gamma_{\beta}\gamma_{\alpha}) \;\; .
\label{1}
\end{equation}

It is convenient to use the recursion relation for three final particles phase
space:
\begin{equation}
d\tau_3 =\frac{1}{2\pi}d\tau_2 (Q,q_1,q_2)d\tau_2 (p_1+k_1, p_2,Q)dQ^2 \;\; ,
\label{2}
\end{equation}
where $Q$ is the sum of the muon momenta $Q=q_1+q_2$ (all momenta are
defined in Fig.3), and to perform the integration of the square of the
amplitude just in the order, shown in (\ref{2}). Leaving precise formulas for
the next Section, here we trace the derivation of the cross section in
the leading logarithmic approximation. After the integration over the
direction of $\mu$-meson momenta which is conveniently performed in the center
of mass system of two muons we come to the following double differential cross
section:
\begin{equation}
\frac{d^2 \sigma}{dQ^2 dq^2}=\frac{\alpha^2\mu^2_{\nu}}{\pi}
\ln\left(\frac{1+v}{1-v}\right)\frac{1}{(Q^2 +q^2)^2}
\left[1-\frac{2Q^2}{Q^2+q^2}+\frac{2Q^4}{(q^2+Q^2)^2}\right] \;\; ,
\label{3}
\end{equation}
where $-q^2 =(p_1 -p_2)^2$ is virtuality of the  photon and
$v=\sqrt{1-\frac{4m_{\mu}^2}{Q^2}}$. Integrating eq.(\ref{3}) over $q^2$
in the domain $0 <q^2 < s-Q^2$, $s \equiv (p_1 +k_1)^2$, $Q^2 \ll s$, we
obtain:
\begin{equation}
\frac{d\sigma}{dQ^2} =\frac{2}{3}\frac{\alpha^2
\mu^2_{\nu}}{\pi Q^2} \ln \left(\frac{1+v}{1-v}\right)\approx
\frac{2}{3}\frac{\alpha^2\mu_{\nu}^2}{\pi Q^2} \ln
\left(\frac{Q^2}{m_{\mu}^2}\right) \label{4}
\end{equation}

Finally, integrating eq.(\ref{4}) over $Q^2$ in the domain $4m_{\mu}^2 <Q^2
<s$ we get with the double logarithmic accuracy:
\begin{equation}
\sigma =\frac{1}{3}\frac{\alpha^2 \mu_{\nu}^2}{\pi}\ln^2
\left(\frac{s}{m_{\mu}^2}\right) \;\; .
\label{5}
\end{equation}

Let us note that the log factor in eq.(\ref{4}) $\ln\frac{1+v}{1-v} \approx
\ln\left(\frac{Q^2}{m_{\mu}^2}\right)$ comes from the integration over directions
of muon momenta and corresponds to $\gamma^*\gamma\to\mu^+\mu^-$ part of the
diagram (lower block in Fig.3), while the second log in eq.(\ref{5}) comes from
the integration over the mass of $\mu^+\mu^-$-pair $Q^2$.

\section{Exact formulas for the reaction \mbox{\boldmath$\gamma\nu\to\mu^+
\mu^- \nu$} }

In this Section precise formulas which describe the process
$\gamma\nu \to \mu^+\mu^- \nu$ shown in Fig.3 will be presented. 
We start from the double
differential cross section $d^2\sigma/dq^2 dQ^2$, where $q^2 \equiv 2p_1 p_2$
equals the minus square of the momentum of a virtual photon and $Q^2 =(q_1
+q_2)^2$ is the invariant mass of $\mu^+ \mu^-$-pair. Let us 
define the following variables: $s=(p_1 +k_1)^2$ is the 
Mandelstam variable; $v=\sqrt{1-\frac{4m_{\mu}^2}{Q^2}}$ is $\mu$-meson
velocity in the $\mu^+ \mu^-$ pair center of mass system (for brevity 
instead of $m_{\mu}$
we will write $m$); finally $\gamma$ is the angle between the momenta of
 the initial neutrino and photon in the coordinate system where
the center of mass of the 
produced $\mu^+ \mu^-$-pair is at rest (and where the integration of their
phase space is straightforward):
\begin{equation}
\cos\gamma = 1-\frac{2Q^2 s}{(Q^2 +2p_1 p_2)(s-2p_1p_2)}
\label{15}
\end{equation}

After all these preliminaries we can present the double differential cross
section:
\begin{eqnarray}
\frac{d^2 \sigma}{dq^2 dQ^2} & = & \frac{2\alpha^2 \mu_{\nu}^2}{\pi s^2 q^2}
\left\{\left\{ \frac{1}{4}\ln\left(\frac{1+v}{1-v}\right)
\left\{\frac{(s-q^2)^2}{(Q^2 +q^2)^2}
[2m^2(-2+2\frac{m^2}{Q^2})(1-\cos^2\gamma) + \right. \right. \right.
\nonumber \\
& + & Q^2(1+\cos\gamma)+q^2\cos\gamma(1-\cos\gamma)+
2\frac{m^2}{Q^2}q^2(1-\cos^2\gamma)]- \nonumber \\
&-&\frac{s(s-q^2)}{(Q^2 +q^2)}(1+\cos\gamma) +  \label{16} \\
& + & \left. 2m^2\frac{(s-q^2)}{(Q^2 +q^2)^2}2s(1+\cos\gamma)+\frac{2q^2}{(Q^2
+q^2)^2}[2(Q^2 +m^2)(s -q^2)+2m^2 q^2] \right\} + \nonumber \\
& + & \frac{v}{2}\frac{(s-q^2)^2}{(Q^2 +q^2)^2} \sin^2\gamma(m^2 +Q^2)
-\frac{Q^2 vs}{2} \frac{(s-q^2)}{(Q^2 +q^2)^2} (1+\cos\gamma) +\frac{1}{4}v
\frac{(s-q^2)^2}{(Q^2 +q^2)} \times \nonumber \\
& \times &  [3(\cos^2\gamma -1) -2(\cos\gamma +1)]
+v\frac{s-q^2}{Q^2 +q^2} [-q^2
+\frac{s}{2}(1+\cos\gamma)- \nonumber \\
&-& \left. \left.\frac{Q^2}{2}\cos^2\gamma\frac{(s-q^2)}{Q^2
+q^2}+\frac{s}{2}] \right\} \right\} \;\; . \nonumber 
\end{eqnarray}

Note that the expression in the double curly brackets 
 is proportional to $q^2$
for $q^2 \to 0$ (because in this limit $\cos\gamma =-1$) and cancels the pole
$1/q^2$. Thus it can be presented in the form:

\begin{eqnarray}
\frac{d^2 \sigma}{dq^2 dQ^2} 
& = & \frac{\alpha^2 \mu_{\nu}^2}{\pi s^2}
\left\{
\ln\left(\frac{1+v}{1-v}\right)
\left\{\frac{1}{t}\left( -s-2m^2\right)+\frac{1}{t^2}\left( s^2+2sQ^2+
2m^2(4s+Q^2)\right)+\right. \right.\nonumber \\
& + & 
\frac{1}{t^3}\left(-2sQ^2(s+Q^2)+2m^2(-4s^2-6sQ^2+4m^2s)\right)+ \label{20}\\
& + & \left.
\frac{1}{t^4}\left(2s^2Q^4+2m^2(6s^2Q^2-4m^2s^2)\right) \right\} + \nonumber \\
& + & 
v
\left\{\frac{1}{t}\left( s+Q^2\right)+\frac{1}{t^2}\left( -(s+Q^2)^2-6sQ^2
\right)+\right. \nonumber \\
& + & \left.\left.
\frac{1}{t^3}\left(8sQ^2(s+Q^2)+4m^2sQ^2\right)+
\frac{1}{t^4}\left(-8s^2Q^4-4m^2s^2Q^2\right) \right\} \right\} \;\;,\nonumber 
\end{eqnarray}

where $t=Q^2+q^2 $ .

Neglecting the terms proportional to $m^2$  and leaving only those terms
which will produce $\log^2 s$ after the integration over $Q^2$ one can get 
equation (\ref{3}) presented in Section 2.

Integrating (\ref{20}) over $q^2$ in the interval $0 < q^2 < s-Q^2$, we
obtain:
\begin{eqnarray}
\frac{d\sigma}{dQ^2} &=& \frac{\alpha^2 \mu_{\nu}^2}{2\pi Q^2} \left\{
\ln(\frac{1+v}{1-v}) \left[\frac{4}{3}+2\frac{Q^2}{s}\ln(\frac{Q^2}{s})
-2\frac{Q^4}{s^2} +\frac{2}{3}\frac{Q^6}{s^3} -\frac{16}{3}\frac{m^4}{Q^4} +
\right. \right. \nonumber \\
&+& \left. 4\frac{m^2}{s}+8\frac{m^4}{sQ^2} +\frac{4m^2 Q^2}{s^2}
\ln(\frac{Q^2}{s}) -\frac{4m^2 Q^2}{s^2} -\frac{8}{3} \frac{m^4 Q^2}{s^3}
\right] + \label{17}\\
&+& v \left[\frac{2}{3} -6\frac{Q^2}{s} +6\frac{Q^4}{s^2}
-\frac{2}{3}\frac{Q^6}{s^3} +2(\frac{Q^2}{s} +\frac{Q^4}{s^2})
\ln(\frac{s}{Q^2}) - \right. \nonumber \\
&-& \left. \left.\frac{4}{3}\frac{m^2}{Q^2}\left(\frac{2s^3 -3Q^2 s^2
+Q^6}{s^3}\right) \right] \right\} \;\; , \nonumber
\end{eqnarray}
and the leading term presented in eq. (\ref{4}) is given by the first term in the
first square brackets.

Finally, integrating (\ref{17}) in the domain $4m^2 < Q^2 < s$ we obtain the 
expression for the total cross section of the reaction $\gamma \nu \to \mu^+
\mu^- \nu$:
\begin{eqnarray}
\sigma & = &
\frac{\alpha^2\mu_\nu^2}{2\pi}\left\{
\left\{\frac{1}{2}\ln(\frac{1+v_m}{1-v_m}) \ln(\frac{s}{m^2})+
F(-\frac{1+v_m}{2})-F(-\frac{1-v_m}{2}) \right\} 
\left[\frac{4}{3}+r+\frac{r^2}{4}\right] + \right. \nonumber \\
& + & \left.
\ln(\frac{1+v_m}{1-v_m})
\left[-\frac{19}{9}+r-\frac{r^2}{4}+\frac{7}{72}r^3\right] + 
v_m\left[\frac{46}{27}+\frac{17}{27}r+\frac{7}{36}r^2\right] \right\} = \\
& = & 
\frac{\alpha^2\mu_\nu^2}{3\pi}\left\{
\ln^2\frac{s}{m^2}-\frac{19}{6} \ln\frac{s}{m^2} +
\frac{23}{9}-\frac{\pi^2}{3}+ O\left(\frac{m^2}{s}\ln^2\frac{s}{m^2}\right)
\right\} \;\;, \nonumber  \label{18} 
\end{eqnarray}

where $r=4m^2/s=1-v_m^2$ and 

\begin{eqnarray}
F(s) = \int^s_0 \ln(1+x)\frac{dx}{x}\;\;.
\label{19} 
\end{eqnarray}

\section{Reaction \mbox{\boldmath$\nu N\to\mu^+\mu^- \nu N$} in the equivalent
photon \\ approximation}

For the spectrum of the invariant mass of $\mu^+ \mu^-$ pair produced in $\nu
N$-scattering we obtain:
\begin{eqnarray}
\frac{d\sigma}{dQ^2}& = &\int\limits^{E_{\nu}}_{Q^2/m_{\mu}}
\left(\frac{d\sigma}{dQ^2}\right)_r \frac{2Z^2\alpha}{\pi} \ln
\left(\frac{E_{\nu}}{\omega}\right) \frac{d\omega}{\omega} = \nonumber \\
&=& \frac{2}{3} \frac{Z^2 \alpha^3 \mu_{\nu}^2}{\pi^2 Q^2} \ln
\frac{Q^2}{m_{\mu}^2} \ln^2 \left(\frac{E_{\nu}m_{\mu}}{Q^2}\right) \;\; ,
\label{6}
\end{eqnarray}
where $\omega$ is the  energy of the photon radiated by the nucleus, 
$E_{\nu}$ is the energy of the initial neutrino and $Z$ is the charge of the
nuclei; $(d\sigma/dQ^2)_r$ corresponds to the photoproduction by a real photon
and is given by eq.(\ref{4}).
Virtuality of the photon emitted by a nucleus varies in the following region:
$(Q^2/E_{\nu})^2 < k^2 < Q^2$. Since in the leading logarithmic
approximation $Q^2 \ll E_{\nu}m_{\mu}$, one can neglect a nuclear form factor
for the experimentally interesting values of neutrino energies, $E_{\nu} \sim
20\div 160$ GeV \cite{1}. Integrating over $\mu^+ \mu^-$ invariant mass from
$m_{\mu}^2$ up to $E_{\nu}m_{\mu}$ we get in the leading logarithmic
approximation:
\begin{equation}
\sigma_{\nu N \to \mu^+ \mu^- \nu N} =
\frac{Z^2\alpha^3}{18\pi^2} \mu_{\nu}^2 \ln^4
\left(\frac{E_{\nu}}{m_{\mu}}\right) \;\; ,
\label{7}
\end{equation}
the  $\log^4 E_{\nu}$ dependence of the cross section announced in the
Abstract.

Surely the same result for the total cross section can be obtained directly from
eq.(\ref{5}). Choosing the coordinate system where the energy of the initial neutrino
is of the order of the mass of the muon, $E^{\prime}_{\nu}  \approx
m_{\mu}$, we obtain:
\begin{eqnarray}
\sigma_{\nu N \to \mu^+ \mu^- \nu_N}& =&\int\limits^{E_{\nu}}_{m_{\mu}}
\sigma_r(s) \frac{2Z^2\alpha}{\pi}\ln \left(\frac{E_{\nu}}{\omega}\right)
\frac{d\omega}{\omega} = 
\label{8}\\
&=&\frac{2Z^2 \alpha^3}{3\pi^2}\mu_{\nu}^2 \int\limits^{E_{\nu}}_{m_{\mu}} \ln
\left(\frac{E_{\nu}}{\omega}\right)\ln^2\left(\frac{\omega}{m_{\mu}}\right)
\frac{d\omega}{\omega} = \frac{Z^2 \alpha^3}{18\pi^2} \mu_{\nu}^2 \ln^4
\left(\frac{E_{\nu}}{m_{\mu}}\right)  \;\; , \nonumber
\end{eqnarray}
which coincides with eq.(\ref{7}) (here $s\approx \omega E^{\prime}_{\nu}
\approx \omega m_{\mu}$).

\section{Comparison with experimental data and bounds on \\ magnetic moment of
muon neutrino}

Production of $\mu^+ \mu^-$-pairs in $\nu_{\mu}(\bar{\nu}_{\mu})$
$N$-scattering observed by experimentalists is well described by the 
Standard Model (Fig.1) \cite{1, 8}. In the CHARM II experiment \cite{1} there
were  used glass target (for the target used in this experiment mean charge
square per nucleus is $<Z^2>=97.6$\footnote{We are grateful to
A.N.Rozanov for this information.}) and neutrino beam
with mean energy 22.3 GeV. Experimental  and theoretical cross sections  are:

\begin{eqnarray}
\sigma_{exp}^{[1]}=[3.0\pm 0.9(stat.)\pm 0.5(syst.)]\times
10^{-41}\;cm^2 \;per\;glass\; nucleus,
\end{eqnarray}
\begin{equation}
\sigma_{SM}^{[1]}=(1.9\pm 0.4)\times 10^{-41}\;cm^2 \;per\;glass\;
nucleus. 
\end{equation}

In the CCFR experiment \cite{8} high energy neutrino beam
with $<E_\nu> =160\;GeV$ was used. Iron (Z=26) was taken as a target. The
number of trident events observed is 
\begin{equation}
N(data)=37.0\pm 12.4\;,
\label{16}
\end{equation}
that corresponds to the following cross section:
\begin{equation}
\sigma_{exp}^{[2]} =(4.7\pm 1.6)E_\nu(GeV)\times 10^{-42}\;cm^2\;per\;
 Fe\; nucleus\;.
 \label{17}
\end{equation}
The number of trident events predicted by the Standard Model is \cite{8}
\begin{equation}
N(trident,\;Standard\;Model)=45.3\pm 2.3\;.
 \label{18}
\end{equation}
From (\ref{16})-(\ref{18}) we obtain:
\begin{equation}
\sigma_{SM}^{[2]} =5.75E_\nu(GeV)\times 10^{-42}\;cm^2\;per\;
 Fe\; nucleus\;.
\end{equation}
As for error in this case, according to \cite{8} it doesn`t exceed $5\%$ 
(see (\ref{18})) that is much smaller than the experimental error in (\ref{17})
and we omit it.

Using these experimental data and approximate formula (\ref{8}) 
 we are able to get restriction on the magnetic moment 
of a neutrino. At 90\% C.L. 
we obtain the following bounds on the muon
neutrino magnetic moment:
\begin{equation}
\mu_{\nu} \la 6.5 \cdot 10^{-8}\mu_B \;\; ,\;\;\mu_B
=\frac{e}{2m_e}\;\;,\;\;e=\sqrt{4\pi\alpha}\;\; ,
\label{9}
\end{equation}
from CHARM II experimental data, and
\begin{equation}
\mu_{\nu} \la 4.0 \cdot 10^{-8}\mu_B 
 \label{10}
\end{equation}
from CCFR experimental data, 
which are approximately two orders
of magnitude weaker than the bound on $\mu_{\nu_{\mu}}$ obtained from
$\nu_{\mu}e$-scattering data \cite{5}.

As it was noted in \cite{6} the neutrino magnetic moment can depend on $q^2$ in
such a way, that it increases for small $q^2 \la  m^2_{\nu_{\mu}}$ and can
reach at most $4\cdot 10^{-6} \mu_B$. However, from (\ref{3}) we see that
the contribution of small $q^2$ into $d\sigma/dQ^2$ is suppressed as
$q^2/Q^2 <(m_{\nu_{\mu}}/m_{\mu})^2 <10^{-6}$ that is why the production of muon
pairs do not allow to study $\mu_{\nu_{\mu}}$ at small $q^2$.

\section{Conclusions}

Our main result is the observation of rapid increase with the energy
of the cross section of the charged lepton pair
production in $\nu N$-collisions in the case of nonzero
neutrino magnetic moment, $\sigma_{\mu} \sim \log^4 E$. Here we will
show how this dependence changes to the famous $\sigma_e \sim \log^3 E$
dependence in the case of two charged particles collision \cite{2, 3}. Unlike
the magnetic moment case the vertex of the photon emission by a charged
particle does not contain the photon momentum $q$. That is why the factor $\mu^2$ in
the spectrum described by eq.(\ref{3}) converts into $1/q^2$. The logarithmic
factor in (\ref{3}) $\log\left(\frac{1+v}{1-v}\right) \approx
\log(Q^2/m_{\mu}^2)$ remains, since it comes from the $\gamma\gamma^*
\to\mu^+\mu^-$ blocks of the diagrams shown in Fig.3 which are universal.
The integral over $q^2$ starts to diverge in the infrared. When $\gamma^*$ is
emitted by a massless neutrino, the values of $q^2$ can reach zero. However,
for the massive electron $q^2$ minimum is above zero:
\begin{equation}
(q^2)_{min} \approx m_e^2 \frac{Q^4}{s(s-Q^2)} \;\; .
\label{11}
\end{equation}

For the spectrum over $Q^2$ we get:

\begin{equation}
\left(\frac{d\sigma}{dQ^2}\right)_{\gamma e \to \mu^+ \mu^- e} \sim \ln
\left(\frac{Q^2}{m_{\mu}^2}\right) \int\limits^{s-Q^2}_{(q^2)_{min}}
\frac{dq^2}{(Q^2 +q^2)^2 q^2} = \frac{1}{Q^4}
\ln \left(\frac{Q^2}{m_{\mu}^2}\right)
\ln \left(\frac{s^2}{m_e^2 Q^2}\right) \;\; .
\label{12}
\end{equation}

Comparing with the analogous spectrum in the magnetic case given by eq. (\ref{4})
we see that now one extra log persists. However integrating over $Q^2$ we
get:
\begin{equation}
\sigma_{\gamma e \to\mu^+ \mu^- e} \sim \frac{1}{m_{\mu}^2} \ln
\left(\frac{s}{m_e m_{\mu}}\right) \;\; ,
\label{13}
\end{equation}
and comparing with eq. (\ref{5}) we note that one log has gone: the double $\log$
behavior changes to one log.

The last step is the conversion of the cross section of $\gamma e \to \mu^+ \mu^- e$
reaction into that of $N e \to N e \mu^+ \mu^-$ with the help of the equivalent
photon (Weiszacker-Williams) approximation:
\begin{eqnarray}
\sigma_{N e \to N e \mu^+\mu^-} &=& \int\sigma_{\gamma e \to \mu^+
\mu^- e}(\omega) \frac{2Z^2\alpha}{\pi} \frac{d\omega}{\omega} \ln
\left(\frac{m_{\mu}}{\omega\frac{m_N}{E_N}}\right) \sim \nonumber \\
& \sim & \frac{1}{m_{\mu}^2}\ln^3 \left(\frac{E_N}{m_N}\right) \;\; ,
\label{14}
\end{eqnarray}
where $s=\omega m_e$. In this way we come to the well-known $\ln^3(E_N/m_N)$
dependence of the cross section of the lepton pair production in the collision of two
charged particles \cite{2,3}.

Finally, let us note that $\log^4E$ behavior of 
the cross section of the reaction $2\to 4$  is  well known in the literature.
In particular, it takes place in the $W^+W^-$ bosons production in the
reaction  $e^+ e^- \to e^+ e^-W^+W^-$ \cite{7}. The reason for such a behavior
is the absence of power decrease with energy of the $\gamma\gamma\to W^+W^-$
cross section; this behaves like a constant unlike that of
$\gamma\gamma\to\mu^+\mu^-$ cross section which behaves like $(\log E^2)/E^2$. 
It is evident that the mechanism of the $\log^4E$ behavior studied in our paper 
is completely different and arises from the photon emission by a neutrino 
magnetic moment.

We are grateful to A.N.Rozanov for inspiration , L.B.Okun for discussion,
and to G.Karl for reading the paper. 
Our investigations are supported by RFBR grants 98-02-17372, 98-02-17453 and
00-15-96562.

\vspace{1cm}

\noindent {\large\bf Figure Captions}

\vspace{0.2cm}

Figure 1: Production of the $\mu^+\mu^-$ pair in $\nu_\mu N$-scattering in the 
Standard Model.

\smallskip

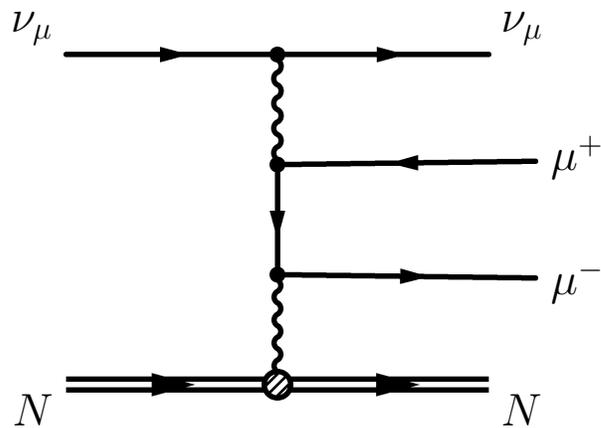
Figure 2: Production of the $\mu^+\mu^-$ pair in $\nu_\mu N$-scattering via 
neutrino magnetic moment.

\smallskip

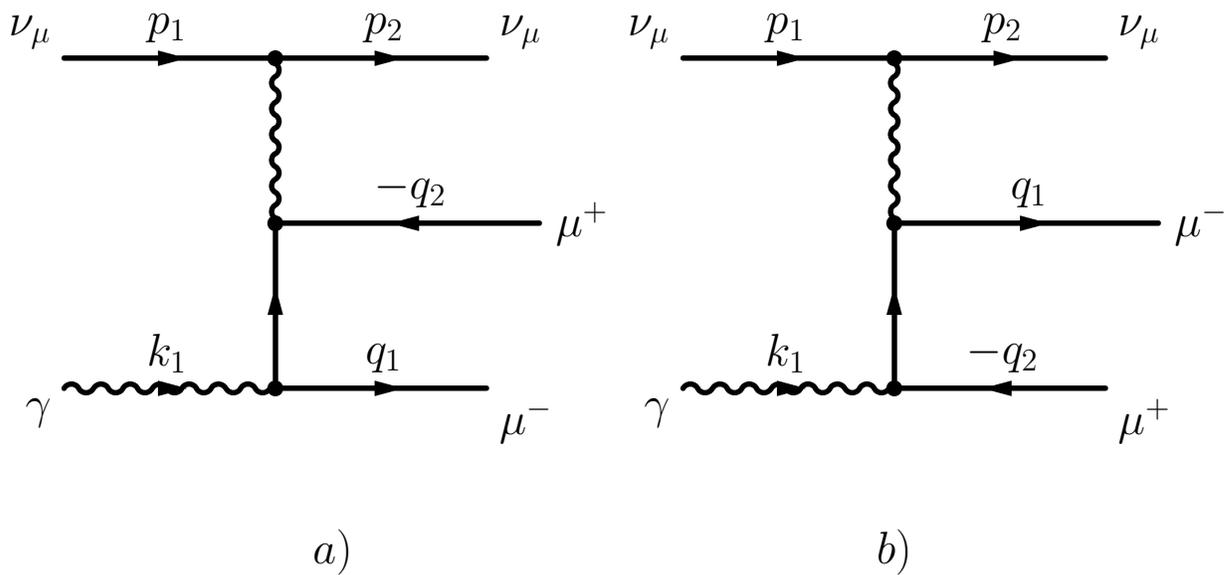
Figure 3: Photoproduction of the $\mu^+\mu^-$ pair on  neutrino via 
neutrino magnetic moment.

\Large

\begin{fmffile}{ts}

$$
\begin{array}[10cm]{cc}
\begin{fmfgraph*}(200,125) \fmfpen{thick}
\fmfleft{i1,i2}
\fmfright{o1,o2,o3,o4} 
\fmf{dbl_plain_arrow}{i1,v4}
\fmf{dbl_plain}{v4,v1}
\fmf{dbl_plain_arrow}{v1,o1}
\fmf{fermion}{i2,v3,o4}
\fmffreeze
\fmf{fermion}{o3,v3,v2,o2}
\fmffreeze
\fmf{photon}{v1,v2}
\fmfdot{v3}
\fmfdot{v2}
\fmfblob{.05w}{v1}
\fmflabel{$ N$}{i1}
\fmflabel{$N$}{o1}
\fmflabel{$\nu_\mu$}{i2}
\fmflabel{$\mu^+$}{o3}
\fmflabel{$\mu^-$}{o2}
\fmflabel{$\nu_\mu$}{o4}
\end{fmfgraph*}\;\;\;\;\;\;\;\;\;\;
&
\begin{fmfgraph*}(200,125) \fmfpen{thick}
\fmfleft{i1,i2}
\fmfright{o1,o2,o3,o4} 
\fmf{dbl_plain_arrow}{i1,v1,o1}
\fmf{fermion}{i2,v3,o4}
\fmffreeze
\fmf{fermion}{v4,v2}
\fmf{photon}{v1,v2}
\fmf{photon}{v3,v4}
\fmffreeze
\fmf{fermion}{o3,v4}
\fmf{fermion}{v2,o2}
\fmfdot{v3}
\fmfdot{v2}
\fmfdot{v4}
\fmfblob{.05w}{v1}
\fmflabel{$ N$}{i1}
\fmflabel{$N$}{o1}
\fmflabel{$\nu_\mu$}{i2}
\fmflabel{$\mu^+$}{o3}
\fmflabel{$\mu^-$}{o2}
\fmflabel{$\nu_\mu$}{o4}
\end{fmfgraph*}\\
&\\
&\\
{\rm Fig. 1 }
& {\rm Fig. 2 }\\
&\\
&\\
&\\
&\\
\begin{fmfgraph*}(200,125) \fmfpen{thick}
\fmfleft{i1,i2}
\fmfright{o1,o2,o3} 
\fmf{photon,label=$k_1$,label.side=left}{i1,v2}
\fmf{fermion,label=$q_1$,label.side=left}{v2,o1}
\fmf{fermion,label=$p_1$,label.side=left}{i2,v3}
\fmf{fermion,label=$p_2$,label.side=left}{v3,o3}
\fmffreeze
\fmf{phantom_arrow}{i1,v2}
\fmf{fermion}{v2,v4}
\fmf{photon}{v3,v4}
\fmffreeze
\fmf{fermion,label=$-q_2$,label.side=right}{o2,v4}
\fmfdot{v3}
\fmfdot{v2}
\fmfdot{v4}
\fmflabel{$\gamma$}{i1}
\fmflabel{$\nu_\mu$}{i2}
\fmflabel{$\mu^+$}{o2}
\fmflabel{$\mu^-$}{o1}
\fmflabel{$\nu_\mu$}{o3}
\end{fmfgraph*}
&
\begin{fmfgraph*}(200,125) \fmfpen{thick}
\fmfleft{i1,i2}
\fmfright{o1,o2,o3} 
\fmf{photon,label=$k_1$,label.side=left}{i1,v2}
\fmf{fermion,label=$-q_2$,label.side=right}{o1,v2}
\fmf{fermion,label=$p_1$,label.side=left}{i2,v3}
\fmf{fermion,label=$p_2$,label.side=left}{v3,o3}
\fmffreeze
\fmf{phantom_arrow}{i1,v2}
\fmf{fermion}{v2,v4}
\fmf{photon}{v3,v4}
\fmffreeze
\fmf{fermion,label=$q_1$,label.side=left}{v4,o2}
\fmfdot{v3}
\fmfdot{v2}
\fmfdot{v4}
\fmflabel{$\gamma$}{i1}
\fmflabel{$\nu_\mu$}{i2}
\fmflabel{$\mu^+$}{o1}
\fmflabel{$\mu^-$}{o2}
\fmflabel{$\nu_\mu$}{o3}
\end{fmfgraph*}\\
&\\
&\\
\;\;\;\;\;\;\;\;\; a) & b) \\
\end{array}
$$
\begin{center}
Fig. 3
\end{center}

\end{fmffile}  



\newpage

\begin{thebibliography}{99}
\bibitem{1} CHARM II Collaboration, Phys. Lett. {\bf 245B} (1990) 271. 
\bibitem{8} CCFR Collaboration, Phys. Rev. Lett. {\bf 66}, 3117 (1991). 
\bibitem{9} NUTEV Collaboration, hep-ex/9811012 (1998).
\bibitem{2} L.D.Landau, E.M.Lifshitz, Phys. Zs. Sowjet, {\bf 6} (1934) 244. 
\bibitem{3} V.M.Budnev, I.F.Ginzburg, G.V.Meledin, and V.G.Serbo, 
Phys. Rep. {\bf 15C} (1975) 181.
\bibitem{5} Particle Data, The European Physical Journal {\bf C3} (1998) 1.
\bibitem{6} J.M.Frere, R.B.Nevzorov, M.I.Vysotsky, Phys. Lett. {\bf B394} 
(1997) 127.
\bibitem{7} O.P.Sushkov, V.V.Flambaum, I.B.Khriplovich, Yad. Fiz. 20 (1974) 
1016.
\end{thebibliography}
\end{document}